\documentclass[epj,twocolumn]{webofc}
\usepackage[varg]{txfonts}   
\usepackage{multirow}

\woctitle{NSRT15}
\begin{document}
\title{Impact of phonon coupling on the radiative nuclear reaction characteristics}

\author{Oleg Achakovskiy\inst{1} \fnsep\thanks{\email{oachakovskiy@ippe.ru}} \and
        Alexander Avdeenkov\inst{1} \and
        Sergei Kamerdzhiev\inst{2}
}

\institute{Institute for Physics and Power Engineering, Obninsk, Russia
\and
           National Research Centre "Kurchatov Institute", Moscow, Russia
          }

\abstract{%
 The pygmy dipole resonance and photon strength functions (PSF) in stable and unstable Ni and  Sn isotopes are  calculated 
within the microscopic self-consistent version of the extended theory of finite fermi systems  in the quasiparticle time blocking approximation.  The approach includes phonon coupling (PC) effects  in addition to the standard QRPA approach. 
 The  Skyrme force SLy4 is used. A pygmy dipole resonance in $^{72}$Ni is predicted at the mean energy of 12.4 MeV exhausting 25.7\% of the total energy-weighted sum rule. With our microscopic E1 PSFs in the EMPIRE 3.1 code,  the following radiative nuclear reaction characteristics  have been calculated  for several stable and unstable even-even Sn and Ni isotopes:  1) neutron capture cross sections, 2) corresponding neutron capture gamma-spectra, 3) average radiative widths of neutron resonances. Here, three variants of the microscopic nuclear level density models have been used and a comparison with the phenomenological generalized superfluid model has been performed. In all the considered properties, including the recent experimental data for PSF in Sn isotopes, the PC contributions turned out to be significant, as compared with the QRPA one,  and necessary to explain the available experimental data. 
}
\maketitle
\section{Introduction}
\label{intro}

The  information about photon strength function (PSF)   is necessary to calculate all characteristics 
of nuclear reactions with gamma-rays, in particular,
the radiative neutron capture cross sections, which are
of great astrophysical \cite{gor} and nuclear engineering \cite{muhab} interest.
Commonly, one parametrizes the PSF phenomenologically using, for example,  generalized Lorentzian models \cite{ripl2,ripl3}.
 The usual definition of PSF contains transitions between excited states. 
For this reason, in order to calculate the PSF, the known Brink-Axel hypothesis is used which states that on each excited 
state it is possible to build a giant dipole resonance (at present, any giant resonance) including it's low-lying part.
In this low-lying energy region, 
there exists the so-called Pygmy-Dipole Resonance (PDR). It exhausts typically about 1-2\% of the Energy Weighted Sum Rule (EWSR) but, nevertheless, it can significantly increase
the radiative neutron capture cross section and affect the nucleosynthesis of neutron-rich nuclei by the r-process \cite{gor}.                                  
  In neutron-rich nuclei, for example, $^{68}$Ni \cite{wiel} and, probably,  $^{72}$Ni, $^{74}$Ni, the EWSR fraction  is much larger. 
Note that for nuclei with small neutron separation energy, less than typically 3--4 MeV, 
the PDR properties are changed  significantly \cite{gor}, and therefore,  phenomenological systematics obtained by fitting  characteristics of stable nuclei cannot be applied. 
Because the Brink-Axel hypothesis, probably, is valid,  the PSF is connected very simply
with the photoabsorption cross section  and, therefore, with the PDR field,
see \cite{ripl2, savran2013, kaev2014}. 
For all these reasons, during the last decade there has been an increasing interest in the investigations of the excitations in the PDR
energy region manifested both in "pure" low-energy nuclear physics \cite{savran2013, paar2007}  and in the nuclear data field \cite{gor, ripl2, ripl3}.
 
 The experiments in the PDR energy region \cite{toft10,toft11,uts2011,tsoneva} have given  additional information about the PDR and PSF structures. The PSF structures at 8--9 MeV in six Sn isotopes obtained  by the Oslo method \cite{toft10,toft11} could not be explained within  the standard phenomenological approach.
 In order  to explain the experiment, it was necessary to add "by hand" some additional strength of about 1--2\% of the EWSR.

Given the importance of PSF both in astrophysics \cite{gor} and nuclear engineering \cite{muhab}, 
microscopic investigations are required, especially when extrapolations to exotic nuclei are needed.
Mean-field approaches using effective nucleon interactions, such as the Hartree-Fock Bogoliubov method and
the quasi-particle random-phase approximation (HFB+QRPA) \cite{gor}, allow systematic self-consistent 
studies of isotope chains, and indeed have been included in modern nuclear reaction codes like EMPIRE \cite{empire} and TALYS \cite{koning12}. Such an approach 
 is  of higher predictive power in comparison with phenomenological models. However, as we discuss below and as confirmed by recent experiments, the  HFB+QRPA  approach is necessary but not sufficient. To be exact, it should be complemented by the effect describing the interaction of single-particle degrees of freedom with the low-lying collective phonon degrees of freedom, known as the  phonon coupling (PC).

 The results \cite{uts2011} directly confirm the necessity to go beyond the HFB+QRPA method because the PSF structures observed in \cite{toft11} could not be explained within the  HFB+QRPA approach.  In particular, the PC effects discussed in Refs.\cite{PhysRep,ave2011,kaev2014} may be at the origin of such an extra strength. Note that the microscopic PSFs which contained  transitions  between the ground  and excited states, i.e. the PSFs values at the energy point near the neutron separation energy,  have already been estimated long ago within the quasiparticle-phonon model approach, which also includes the phonon coupling \cite{vg}. 
 
 In this work, we use   the self-consistent version of the extended theory of finite fermi systems (ETFFS) \cite{PhysRep} in the quasi-particle time blocking approximation (QTBA) \cite{tselyaev}.  
Our ETFFS (QTBA) method, or simply QTBA,  includes self-consistently the QRPA and PC effects and the single-particle continuum in a discrete form. 
Details of the method are described in Ref. \cite{ave2011}. The method allows us 
to investigate the impact of the PC on nuclear reaction in both stable and unstable  nuclei.  We calculate  the microscopic PSFs in several Sn and Ni  isotopes and use them in the EMPIRE   code to estimate  the  neutron capture cross sections,  corresponding capture gamma-ray spectre and average radiative widths. 

\section{PSFs}
\label{sec-1}

 To calculate the strength function $S(\omega) = dB(E1)/dE$ \cite{PhysRep,ave2011},
which is connected with  the PSF $f(E1)$  as  
$f(E1,\omega)[MeV^{-3}] = 3.487\cdot10^{-7} S(\omega)[fm^2MeV^{-1}]$, 
   we use the well-known SLy4 Skyrme force \cite{chabanat}.  
The ground state is calculated within the  HFB method using the spherical code HFBRAD \cite{bennaceur}. The residual interaction for the (Q)RPA and  QTBA calculations  is derived as the second derivative of the Skyrme functional. 
In all our calculations  we use  a smoothing parameter of 200 keV which effectively accounts for correlations beyond the considered PC 
  which do not show a strong energy dependence.
Such a choice guarantees the correct description of all three characteristics of giant resonances, including their  widths \cite{PhysRep} 
  and, what is important here, this value approximately coincides with the experimental resolution of the Oslo method \cite{toft11}.
 
\begin{figure}

\centering
\includegraphics[width=8cm,clip]{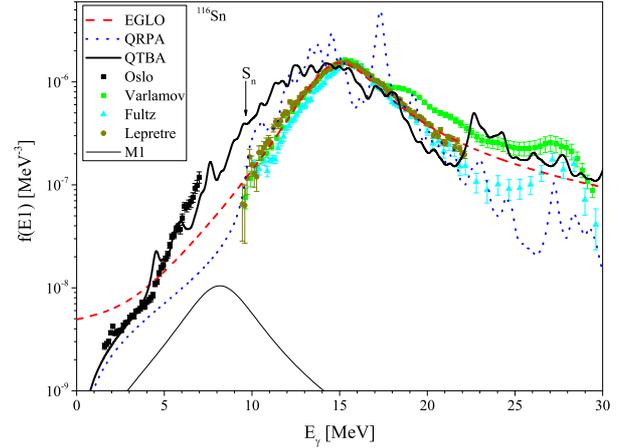}
\caption{(Color online) The E1 photon strength functions for $^{116}$Sn. The dashed lines are obtained within the
phenomenological variant of the EGLO \cite{ripl2}, the dotted line is the QRPA calculation, and the solid line is the
QTBA calculation (the complete microscopic calculation). The arrow marks the neutron separation energy. The
experimental data are taken from \cite{toft10,Varlamov,Fultz,Lepretre}.} 
\label{fig-1}       
\end{figure}

\begin{figure}
\centering
\includegraphics[width=8cm,clip]{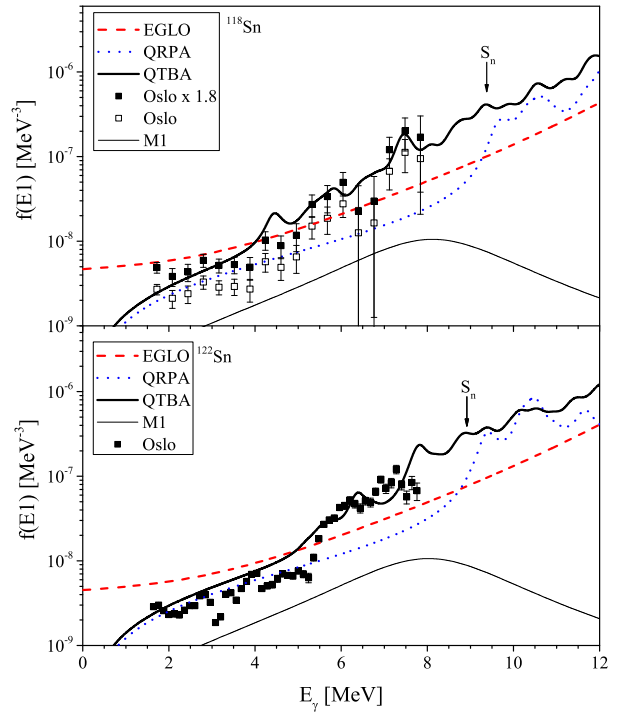}
\caption{The same as for Fig. \ref{fig-1}, but for $^{118}$Sn and $^{122}$Sn and for the PDR energy region. The
experimental data are taken from \cite{toft10, toft11}.} 

\label{fig-2}       
\end{figure} 
 
Figure \ref{fig-1},\ref{fig-2} shows the radiative strength functions for the $^{116,118,122}$Sn isotopes
calculated for three variants of the radiative strength
function: the phenomenological enhanced generalized Lorentzian (EGLO) \cite{ripl2}, as well as the microscopic QRPA and QTBA in each nucleus, in comparison to the existing experimental data for these nuclei.

We obtained:

(i) In contrast to phenomenological approaches, all nuclei exhibit structures caused by the
effects of the QRPA and the effects of phonon coupling. In this case, the difference of the QTBA from
the QRPA, i.e., the contribution from phonon coupling, becomes noticeable at energies E < 10 MeV.
More precisely, structures at about E > 5 MeV are due
only to the effect of phonon coupling. This corresponds to the mentioned experimental data \cite{toft10,toft11}.

(ii) As  expected, the EGLO curves are in general agreement with experimental data for energies above
the neutron threshold, because the EGLO phenomenology was chosen from the corresponding experiments. As  it can be seen in Fig. \ref{fig-1} for $^{116}$Sn, our calculations in the
giant dipole resonance (GDR) energy range  are in worse agreement with the experiment than those in the energy range up to the
binding energy. One of the main reasons for
this is that the smoothing parameter for energies above the nucleon separation energy should depend on the
energy. However, as was shown in \cite{myISINN22_1}, all observed  integral characteristics, including the most important of
them -- the width of the GDR -- are described satisfactorily with a parameter of 200 keV used  here.
 
\begin{figure}

\centering
\includegraphics[width=8cm,clip]{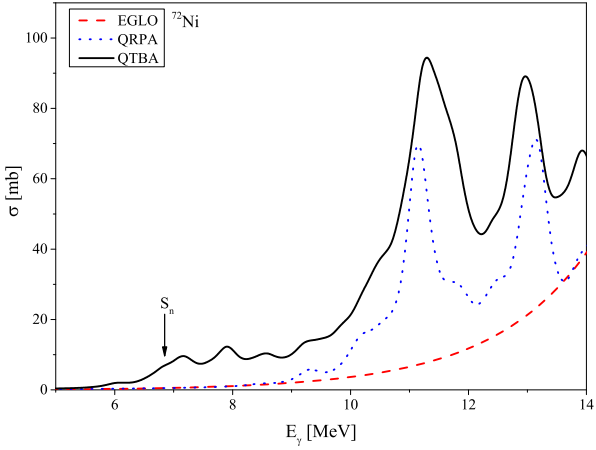}
\caption{The calculated photoabsorption cross section in $^{72}$Ni. See text for details.}
\label{fig-3}       
\end{figure}

In Ref.\cite{wiel}, the PDR in the unstable $^{68}$Ni  nucleus  was measured. It was found that  the PDR is in the range of 7--13 MeV, has a
maximum at an energy of 11 MeV, and exhausts about 5\% of the EWSR. The neutron separation energy is 7.8 MeV, i.e. the PDR in this nucleus is located noticeably above the neutron emission
threshold. The situation in the $^{72}$Ni nucleus should be similar. Predictions for this nucleus are of interest because
of the possibility of the corresponding experiments.
The results of the calculations for the PDR in three nickel isotopes are presented in Table \ref{tab-1} in comparison to the
results for the stable $^{58}$Ni nucleus and shown in Fig. \ref{fig-3} for the PDR in  $^{72}$Ni. 

In Table \ref{tab-1}, the integral parameters of the PDR  are given for three PSF models, i.e. the phenomenological
EGLO, our microscopic QRPA and QTBA
(QRPA+PC). To compare, the 6 MeV interval, where the
PDR was observed in $^{68}$Ni, is considered. In this interval,
the PDR characteristics have been approximated, as usual,
 using  a Lorentz curve by fitting the three moments
of the Lorentzian and theoretical curves \cite{PhysRep}. A reasonable
agreement with experimental data \cite{wiel} for $^{68}$Ni is obtained.
Earlier, a similar calculation was performed for $^{68}$Ni \cite{lrt2010}
using the relativistic QTBA, with two phonon contributions
additionally taken into account. Concomitantly, the
PDR characteristics in $^{72}$Ni have been estimated leading in
this interval to a mean energy of 12.4 MeV and the large
strength of 25.7\% of the total EWSR. In all three isotopes, a
large PC contribution to the PDR strength has been found.

\begin{table}[ht]
\centering
\caption{Integral characteristics of the PDR (mean  energy $E$ in MeV and fraction of the EWSR) in Ni isotopes 
calculated in the (8-14) MeV interval for $^{58}$Ni, $^{72}$Ni and (7--13) MeV interval for $^{68}$Ni (see text for details).}
\label{tab-1}
\begin {tabular}{ l l l l l l}
\hline
\hline
\multirow{2}{*}{Nuclei}&\multicolumn{2}{l}{QRPA}&\multicolumn{2}{l}{QTBA}\\
\cline{2-5}
&$E$&\%&$E$&\%\\
\hline
$^{58}$Ni&13.3&6.0&14.0&11.7\\
$^{68}$Ni&11.0&4.9&10.8&8.7\\
$^{72}$Ni&12.4&14.7&12.4&25.7\\
\hline
\hline
\end{tabular}
 \end{table}

\section{Neutron radiative capture cross sections}
\label{sec-2}

\begin{figure}
\centering
\includegraphics[width=8cm,clip]{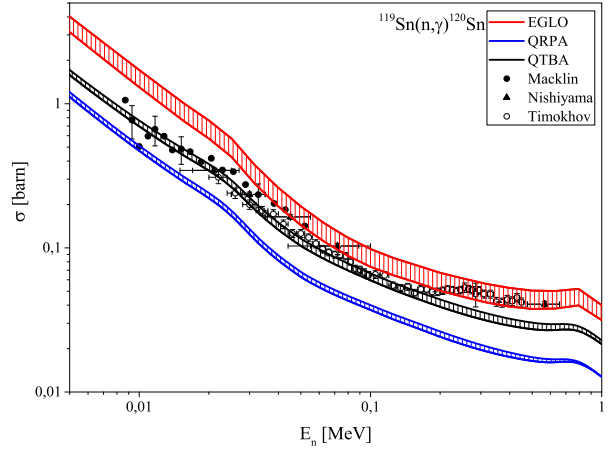}
\caption{$^{119}$Sn(n,$\gamma$) cross section calculated with the EGLO (red), QRPA (blue) and QTBA (black) PSF. The uncertainty bands depict the
uncertainties affecting the nuclear level density predictions \cite{gor08,kon08,hil12}. Experimental data are taken from Refs. \cite{Macklin,Nishi,timokhov}.}
\label{fig-4}       
\end{figure}

As an example,  the  neutron radiative capture cross sections
for  $^{119}$Sn obtained with the QRPA and QRPA+PC photon E1 strength functions are shown in Fig. \ref{fig-4}. 
We also show the uncertainty bands obtained when considering various nuclear level density (NLD) models \cite{gor08,kon08,hil12}. 
One can clearly see that the agreement with experiment up to 0.2 MeV is only
possible when the PC is taken into account. The use of the EGLO PSF with the microscopic NLD model gives an agreement with experiment at higher neutron energies. The details see also in \cite{myPRC2015}.

\section{Capture gamma-ray spectra}
\label{sec-4}

\begin{figure*}
\centering
\includegraphics[width=15cm,clip]{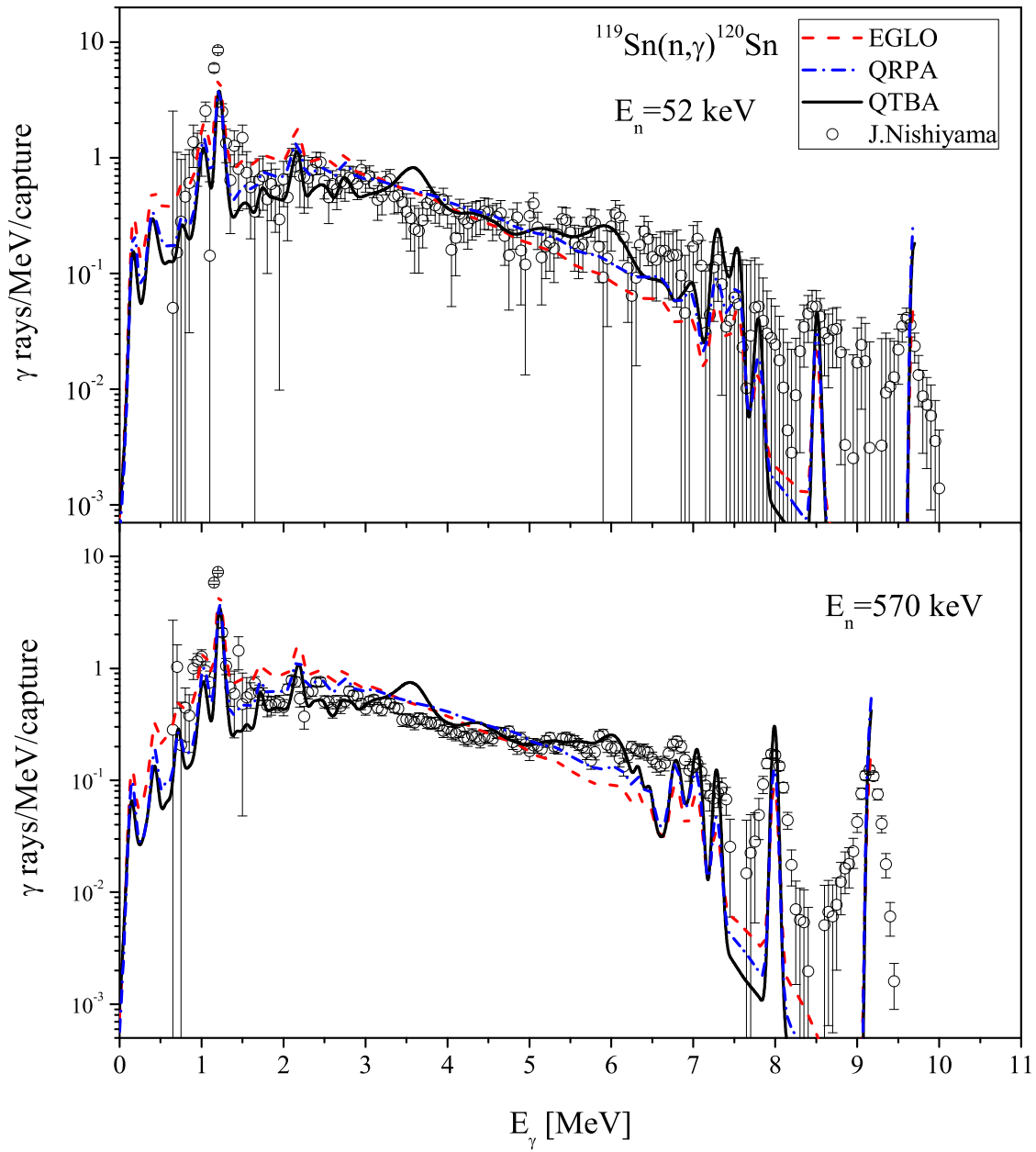}
\caption{Gamma-ray spectra from $^{119}$Sn(n,$\gamma$) for the neutron energy of 52 keV and 570 keV. The microscopic
HFB+combinatorial NLD model \cite{gor08} has been used. Experimental data was taken from \cite{Nishi}.}
\label{fig-5}       
\end{figure*}

The corresponding capture gamma-ray spectra calculated
for the neutron energies of 52 keV and 570 keV are given in Fig. \ref{fig-5}
and compared with with the experimental data \cite{Nishi} and the results obtained with microscopic and phenomenological  EGLO PSF models. Here the microscopic
HFB+combinatorial NLD model \cite{gor08} of nuclear level densities is adopted. 
As compared with  the phenomenological GSM NLD model, which was  used in \cite{myISINN22_2}, the agreement with experiment is better.
Our results show that the PC contribution is significant. For all three PSF variants some structures have been  found. 

\begin{figure}
\centering
\includegraphics[width=8cm,clip]{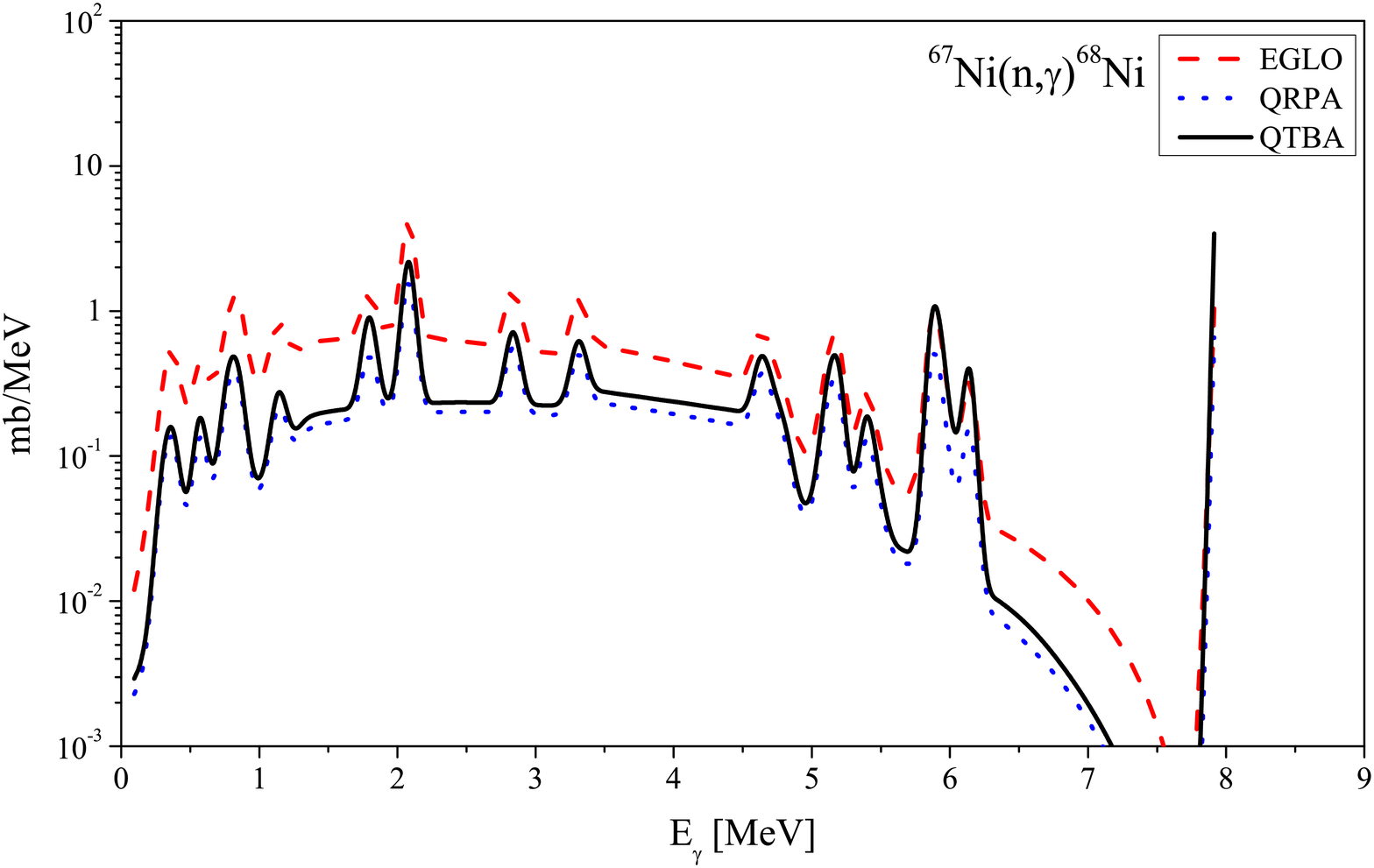}
\caption{Gamma-ray spectra from  $^{68}Ni(n,\gamma$) for the neutron energy of 100 keV. The microscopic
HFB+combinatorial NLD model \cite{gor08} has been used.}
\label{fig-6}       
\end{figure}

We have performed the same calculations for the unstable $^{68}$Ni for the neutron energy 100 keV (see Fig. \ref{fig-6}).
 Here one can see a large difference between results with our two  microscopic  and the phenomenological EGLO  PSF models
which confirms the necessity of using the microscopic approach for unstable nuclei. The similar situation is for  the GSM NLD model in Ref. \cite{myISINN22_2}.

\section{Average radiative widths}
\label{sec-5}

\begin{table*}[ht]
\centering
\caption{Average radiative widths $\Gamma_{\gamma}$ (meV) for s-wave neutrons. For each approach (EGLO, QRPA and QTBA) two NLD models are considered: the phenomenological GSM \cite{ripl2} (first lines) and the microscopic HFB plus combinatorial model \cite{gor08} (second lines). See text for details.}
\label{tab-2}
\renewcommand{\tabcolsep}{0.1cm}
\begin {tabular}{ l l l l l l l l l l l l l l}
\hline
\hline
&$^{110}$Sn&$^{112}$Sn&$^{116}$Sn&$^{118}$Sn&$^{120}$Sn&$^{122}$Sn&$^{124}$Sn&$^{136}$Sn&$^{58}$Ni&$^{60}$Ni&$^{62}$Ni&$^{68}$Ni&$^{72}$Ni\\
\hline
\multirow{2}{*}{EGLO}&147.4&105.5&72.9&46.6&55.0&56.6&49.9&11.1&1096&474&794&166&134\\
&207.9&160.3&108.9&106.7&124.3&110.2&128.7&295.0&2017&1882&1841&982.2&86.4\\
\multirow{2}{*}{QRPA}&45.6&34.4&30.4&22.1&23.8&27.9&22.3&11.2&358&594&623&75.4&83.8\\
                     &71.0&49.7&44.3&40.3&43.0&50.1&68.9&447.8&450.8&1646&490.9&406.4&46.7\\
\multirow{2}{*}{QTBA}&93.5&65.7&46.8&33.1&34.1&35.8&27.9&12.3&1141&971&1370&392&154\\
                     &119.9&87.0&58.4&58.1&61.5&64.0&84.8&509.2&1264&2800&2117&2330&53.8\\
\multirow{2}{*}{Exp.} \cite{muhab}&&&&117 (20)&100 (16)&&&&&2200 (700)&2000 (300)&&\\
$\qquad$ \cite{ripl2}&&&&80 (20)&&&&&&&2200 (700)&&\\
\multirow{2}{*}{M1}&13.0&9.6&8.9&6.1&6.6&7.3&4.9&1.3&46.1&32&23.2&36.0&49.6\\
&29.1&18.1&18.5&13.2&13.4&13.1&15.5&87.2&17.0&52&31.8&81.6&27.5\\

System.&112&109&107&106&105&104&103&73&2650&1900&1300&420&320\\
\hline
\end{tabular}
 \end{table*}

Average radiative widths of neutron resonances $\Gamma_{\gamma}$ are very important properties of  gamma-decay from high-energy nuclear states; they  are used 
for calculations of radiative capture cross sections and other
reactions with gamma-rays.
 There are a lot of  experimental data  for $\Gamma_{\gamma}$ \cite{ripl1,muhab}. For 13 Sn and Ni isotopes, we have calculated with EMPIRE 3.1 the values $\Gamma_{\gamma}$
with the EGLO and our QRPA and QTBA PSF models using  the Generalised Superfluid Model (GSM) NLD \cite{ripl2} and the microscopic HFB plus combinatorial model \cite{gor08}. The predictions are compared in Table \ref{tab-2} with experimental data \cite{muhab}, whenever available, and systematics \cite{ripl1}.   We have found that
   the  PC in stable nuclei  increases  the QRPA contribution  in  the direction of the  systematics and, except for $^{122}$Sn and  $^{124}$Sn, where the increase is limited, the PC leads to an enhancement of  about  50 to 200\%.

 Our  $\Gamma_{\gamma}$  results for $^{118}$Sn, $^{120}$Sn,$^{60}$Ni and $^{62}$Ni, for which  experimental data (not systematics) exists,  are of special  interest.
On the basis of the QTBA strength and the microscopic HFB plus combinatorial NLD \cite{gor08}, we obtain a good  agreement with experiment for $^{60}$Ni, $^{62}$Ni, and reasonable for $^{118}$Sn and $^{120}$Sn. Note that on top of the E1 strength, an M1 contribution following the recommendation of Ref.~\cite{ripl3} is included in the calculation of $\Gamma_{\gamma}$. 
The M1 resonance contribution to $\Gamma_{\gamma}$ has been estimated using the GSM NLD model and the standard Lorentzian parametrization \cite{ripl3} with a width $\Gamma = 4$~MeV (note that such a large $\Gamma$ value is open to question, as discussed in Ref. ~\cite{kaev2006}). 
Such a contribution is found to be of the order of (10-12)\% of the values in the first line of Table ~\ref{tab-2} for Sn isotopes and 4\%, 3\%, 22\% and 16\%  for $^{58}$Ni, $^{62}$Ni, $^{68}$Ni and $^{72}$Ni, respectively.
The agreement  of the $\Gamma_{\gamma}$ values with experiment is found to deteriorate if use is made of the EGLO or QRPA strengths, but also of the GSM NLD. One can also see  that for stable nuclei, the combinatorial NLD model results  are  in a better agreement with the systematics \cite{ripl2} than those obtained with the GSM model.  As far as the EGLO model is concerned, we see that similar conclusions can be drawn. 

\section{Conclusion}
\label{sec-6}
The characteristics of nuclear reactions with gamma-rays
have been calculated within the microscopic self-consistent
approach which takes into account the QRPA and PC effects and uses the SLy4 Skyrme force. Such
a self-consistent approach is of particular relevance for nuclear astrophysics. 
A reasonable agreement with available experimental data has been obtained
thanks to PC. We predict the PDR in
the spherical $^{72}$Ni nucleus at 12.4 MeV with a very large
strength corresponding to 25.7\% of the EWSR. For the
first time, the average radiative widths have been calculated
microscopically with the PC taken into account. In
all the considered quantities, the contribution of PC turned
out to be significant. These results confirm the necessity
of including the PC effects into the theory of nuclear
data both for stable and unstable nuclei.  The phenomenological GSM NLD model used in EMPIRE 3.1 gives, on the whole, the 
worse results than the microscopic HFB plus combinatorial NLD model.

The authors (O.A. and S.K.) acknowledge  Organizing Committee of the NSRT15 conference for the support of their participation in the conference.


\begin{thebibliography}{}

\bibitem{gor}
S. Goriely, E. Khan, V. Samyn, Nucl. Phys. A \textbf{739}, 331 (2004).

\bibitem{muhab}
S.F. Mughabghab, \textit{Atlas of neutron Resonances, Resonance Parameters and Thermal Cross Sections Z=1-100} (Elsevier, Amsterdam, 2006).

 \bibitem{ripl2}
T. Belgya,  O. Bersillon,  R. Capote \textit{et al.},  
 {\it Handbook for Calculations of Nuclear Reaction Data,  RIPL-2},  IAEA-TECDOC-1506 (IAEA,  Vienna,  2006) [http://www-nds.iaea.org/RIPL-2/]. 
 
 \bibitem{ripl3}
 R. Capote, M. Herman, P. Oblozinsky  \textit{et al.}, 
Nuclear Data Sheets {\bf 110},  3107  (2009). See also \emph{https://www-nds.iaea.org/RIPL-3}.

\bibitem{wiel}
 O. Wieland, A. Bracco, F. Camera \textit{et al.}, Phys. Rev. Lett. \textbf{102}, 092502 (2009).

\bibitem{savran2013} 
D. Savran,  T. Aumann,  A. Zilges,  Prog. Part. and Nucl. Phys. {\bf 70},  210  (2013).
 
  \bibitem{kaev2014}
S.P. Kamerdzhiev, A.V. Avdeenkov, O.I. Achakovskiy, Phys. Atom. Nucl. \textbf{77}, 1303 (2014).

 \bibitem{paar2007}
 N. Paar,  D. Vretenar,  E. Khan,  G. Colo,  Rep. Prog. Phys. \textbf{70},  691  (2007).
 
\bibitem{toft10} 
H.K. Toft, A. C. Larsen, U. Agvaanluvsan \textit{et al.}, Phys. Rev. C81, 064311 (2010).
 
  \bibitem{toft11} 
 H.K. Toft, A. C. Larsen, A. B\"{u}rger \textit{et al.}, Phys. Rev. C \textbf{83}, 044320 (2011).
 

 \bibitem{tsoneva}
 R. Schwengner, R. Massarczyk, G. Rusev \textit{et al.}, Phys. Rev. C \textbf{87}, 024306 (2013). 
    
 \bibitem{uts2011}
H. Utsunomiya, S. Goriely, M. Kamata  \textit{ et al.}, Phys. Rev. C \textbf{84}, 055805 (2011).

  \bibitem{empire}  
M. Herman, R. Capote, B.V. Carlson \textit{et al.},  Nucl. Data Sheets,  \textbf{108}  (2007) 2655-2715. See also \emph{http://www.nndc.bnl.gov/empire/index.html}.
 
\bibitem{koning12} 
A.J. Koning and D. Rochman, Nuclear Data Sheets {\bf 113}, 2841 (2012).

 
\bibitem{PhysRep}
 S.Kamerdzhiev, J. Speth, G. Tertychny, Phys. Rep. \textbf{393}, 1 (2004).  

\bibitem{ave2011}
A. Avdeenkov, S. Goriely, S. Kamerdzhiev, S. Krewald, Phys. Rev. C \textbf{83}, 064316 (2011).

\bibitem{vg}
V.G. Soloviev, Ch. Stoyanov and V.V. Voronov, Nucl. Phys.\textbf{ A304}, 503 (1978).

\bibitem{tselyaev}
V. Tselyaev, Phys. Rev. C \textbf{75}, 024306 (2007).

\bibitem{chabanat}
E. Chabanat, P. Bonche, P. Haensel, Nucl. Phys. A \textbf{635}, 231 (1998). 

\bibitem{bennaceur}
K.Bennaceur and J. Dobaczewski, Comp. Phys. Comm, \textbf{168}, 96 (2005).

\bibitem{Varlamov}
V. V. Varlamov, N. N. Peskov, D. S. Rudenko, and M. E. Stepanov, Vopr. At. Nauki Tekh., Ser. Yad. Konstanty \textbf{1}, 2 (2003).

\bibitem{Fultz}
S. C. Fultz, B. L. Berman, J. T. Coldwell \textit{ et al.}, Phys. Rev. \textbf{186}, 1255 (1969).

\bibitem{Lepretre}
A. Lepr\^{e}tre, H. Beil, R. Bergere \textit{ et al.}, Nucl. Phys. A \textbf{219}, 39 (1974).

\bibitem{myISINN22_1}
O.I. Achakovskiy, A.V. Avdeenkov, and S.P. Kamerdzhiev, in Proceedings of the International Seminar on Interaction
of Nuclei with Nucleons, ISINN22 (Dubna, May 27–30, 2014), p. 213.

 \bibitem{lrt2010}
E. Litvinova, P. Ring, V. Tselyaev, Phys. Rev. Lett. \textbf{105}, 022502 (2010).

\bibitem{gor08}
  S. Goriely, S. Hilaire, A.J. Koning, Phys. Rev. C {\bf 78}, 064307  (2008).
  
\bibitem{kon08} 
 A.J. Koning,  S. Hilaire, S. Goriely, Nucl. Phys. A {\bf 810}, 13 (2008).


\bibitem{hil12}
 S. Hilaire,  M. Girod, S. Goriely and A.J. Koning, Phys. Rev. C {\bf 86}, 064317 (2012).
 

\bibitem{Macklin}
R.L. Macklin, T. Inada, J.H. Gibbons,  Washington AEC Office Reports, No.1041, p.30 (1962).

\bibitem{Nishi}
J. Nishiyama, M. Igashira, T. Ohsaki \textit{et al.}, J. Nucl. Sci. Technol. (Tokyo)  \textbf{45}, 352 (2008) 
 
\bibitem{timokhov}
V.M. Timokhov \textit{et al.}, Fiz.-Energ Institut, Obninsk Reports No.1921 (1988). 
 

 \bibitem{myPRC2015}
 O. Achakovskiy, A. Avdeenkov, S. Goriely \textit{et al.}, Phys. Rev. C \textbf{91}, 034620 (2015).

\bibitem{myISINN22_2}
O.I. Achakovskiy, A.V. Avdeenkov, S.P. Kamerdzhiev and D.A. Voitenkov, in Proceedings of the International Seminar on Interaction
of Nuclei with Nucleons, ISINN22 (Dubna, May 27–30, 2014), p. 207.

\bibitem{ripl1}
{\it Reference Input Parameter Library, RIPL-1}, IAEA-TECDOC-1034 (1998).

\bibitem{kaev2006}
S.P. Kamerdzhiev, S.F. Kovalev, Phys. Atom. Nucl. \textbf{69}, 418 (2006).


\end{thebibliography}
\end{document}